\documentclass[12pt]{article}

\usepackage{latexsym,amsmath,amsfonts,amssymb,dsfont}

\newcommand{\newsection}{    
\setcounter{equation}{0}\section}

\def\be{\begin{equation}}
\def\ee{\end{equation}}
\def\bd{\begin{displaymath}}
\def\ed{\end{displaymath}}

\def\ba{\begin{eqnarray}}
\def\ea{\end{eqnarray}}

\def\W{\ensuremath{{\cal W}_\alpha}}
\def\n1{${\cal N}=1$}
\def\q{\ensuremath{\bar{{\cal Q}}^{\dot{\alpha}}}}
\def\tr{\ensuremath{\mathrm {Tr\:}}}
\def\c={\ensuremath{\overset{c.r.}{=}}}
\def\k{\ensuremath{\frac{{\cal W}_{\alpha}^{(k)}{\cal W}^{(k)\alpha}}{z-\Phi_{(k)}}}}
\def\pk1{\ensuremath{\frac{{\cal W}_{\alpha}^{(k+1)}{\cal W}^{(k+1)\alpha}}{z-\Phi_{(k+1)}}}}
\def\p1{\ensuremath{\frac{{\cal W}_{\alpha}^{(1)}{\cal W}^{(1)\alpha}}{z-\Phi_{(1)}}}}
\def\pn{\ensuremath{\frac{{\cal W}_{\alpha}^{(n)}{\cal W}^{(n)\alpha}}{z-\Phi_{(n)}}}}
\def\fr{\ensuremath{\frac{1}{32\pi^2}}}
\def\pW2{\ensuremath{\frac{{\cal W}_{\alpha}^{(2)}{\cal W}^{(2)\alpha}}{z-\Phi_{(2)}}}}

\begin{document}
\font\cmss=cmss10 \font\cmsss=cmss10 at 7pt
\vspace{18pt}
\par\hfill Bicocca-FT-03-9
\vskip .1in \hfill hep-th/0304123

\vspace{1,5cm}
\begin{center}
{\LARGE \textbf{Quivers via anomaly chains}}
\end{center}

\vspace{30pt}

\begin{center}
{\textsl{Roberto Casero and Enrico Trincherini}}

\vspace{20pt}

\textit{Dipartimento di Fisica\\Universit\`{a} di Milano-Bicocca\\
Piazza della Scienza, 3\\
20126 Milano, Italy.}

\end{center}

\vspace{30pt}

\begin{center}
\textbf{Abstract }
\end{center}

\vspace{4pt} {\small \noindent We study quivers in the context of matrix models. We introduce chains of generalized Konishi anomalies to write the quadratic and cubic equations that constrain the resolvents of general affine $\hat{A}_{n-1}$ and non-affine $A_n$ quiver gauge theories, and give a procedure to calculate all higher-order relations. For these theories we also evaluate, as functions of the resolvents, VEV's of chiral operators with two and four bifundamental insertions. As an example of the general procedure we explicitly consider the two simplest quivers $A_2$ and $\hat{A}_1$, obtaining in the first case a cubic algebraic curve, and for the affine theory the same equation as that of $U(N)$ theories with adjoint matter, successfully reproducing the RG cascade result.}  \vfill

\vskip 5.mm
 \hrule width 5.cm
\vskip 2.mm
{\small
\noindent
roberto.casero@mib.infn.it\\enrico.trincherini@mib.infn.it}

\thispagestyle{empty}
\eject

\setcounter{page}{1}
\newsection{Introduction}
Last year's paper \cite{pertwin} by R. Dijkgraaf and C. Vafa raised new interest in the low-energy dynamics of ${\cal N}=1$ 
supersymmetric (chiral) gauge theories. Starting from ideas developed in 
\cite{Bershadsky:1993cx}-\cite{Dijkgraaf:2002vw}, it was conjectured by the 
two authors that the effective superpotential and gauge coupling constants of 
a wide class of ${\cal N} =1$ supersymmetric gauge field theories can be
 calculated via a related matrix model perturbative computation. In this 
correspondence the gauge theory superpotential is translated into an ordinary
 potential for the matrix model. Quite amazingly, all relevant computations
 on the matrix side, are performed in the planar limit, even though no large 
$N$ limit is taken on the gauge theory side.

In the case of $\,U(N)$ gauge theories with matter in the adjoint representation, this same problem was later addressed from a complementary point of view in \cite{CDSW}. By generalizing the Konishi anomaly \cite{Konishi}, the authors of this paper write Ward identities involving some complex functions, which can be regarded as ``generalized'' generating functionals for the vacuum expectation values of chiral superfields. Quite remarkably, it is found that the two approaches are actually equivalent: the generating functional of superfield strength bilinears is identified with the resolvent in the matrix model, and the Ward identities written in the gauge theory are translated in the loop equations of the matrix model.

It seemed a mystery, though, what role the other generating functionals played. It was later found \cite{CSW} \cite{CSW2} that the only other bosonic functional is necessary to go from an off-shell to the on-shell description of the theory. In fact the resolvent equations by themselves are not enough to determine the whole low-energy dynamics. They allow to write the effective superpotential as a function of the glueball superfields, but bring no information about its extremization. The quantum vacua of the theory can be identified by putting conditions on the residues of the second bosonic functional. Some non-trivial questions arise out of this analysis, and more insights on interesting physics will surely come from a deeper study of more general supersymmetric theories from this perspective.

In the last few months both approaches of \cite{pertwin} and \cite{CDSW} have been extended to a wider class of theories, with matter in the fundamental \cite{Seiberg}, bifundamental representations, $SO(N)$ and $Sp(N)$ gauge groups and non-commutative theories. This will also be the line on which our work moves.

We mean to address the study of quiver gauge theories \cite{quivers} using generalized Konishi anomalies. The interest in this problem comes at least from two points. First of all quiver theories are worth being studied by themselves, since they are the low-energy description of D-branes placed at orbifold singularities \cite{quivers}. Moreover, bifundamentals have been considered only from the matrix models point of view \cite{Lazaroiu} \cite{DVdec}, but nothing had been done before using the purely field theoretical approach of \cite{CDSW}.

It was showed in \cite{Klemm} and \cite{mtheorywitten} that ${\cal N}=2$ non-affine quiver theories $A_n$ with $n$ nodes are characterized by an $(n+1)^{th}$ degree Seiberg-Witten curve. When the supersymmetry is broken to ${\cal N}=1$ by turning on superpotentials for the chiral fields in the adjoint, we expect the curve to become a $(n+1)^{th}$ degree algebraic curve, with coefficients proportional to the superpotentials \cite{CKV} \cite{tatar}. Such a curve was obtained in \cite{Lazaroiu} using matrix models for the simplest case $A_2$.

In this paper we will consider both affine and non-affine quivers,  with a generic number $n$ of nodes. For these theories we write the quadratic and cubic constraints the resolvents must obey, and sketch a general procedure to obtain all higher-degree equations. We will show that the constraints we obtain for the $A_2$ theory exactly reproduce the same curve as \cite{Lazaroiu}, and suggest that the algebraic curves of $n>2$ quivers are obtained  as combinations of the higher-order equations we study and are $(n+1)^{th}$ degree polynomials.

The affine $\hat{A}_1$ quiver has been studied using the matrix approach in \cite{DVdec}. We consider it from a gauge theory point of view. We explicitly write the quadratic curve, show that it is the same as the one for a $U(N_1-N_2)$ theory with matter in the adjoint, and that the cubic equation doesn't put any more constraints on the resolvents. We claim this will be the case for all higher-degree equations. This is a successful check of this method since it reproduces what we actually expected: the theory goes through an RG cascade toward an IR fixed point \cite{KS}.

The paper is organized as follows. In the rest of the introduction we give a quick review of quiver gauge theories, reporting all the facts and results which will be related or relevant to our work. In section 2 we give general rules for the chiral ring of operators in the adjoint and bifundamental representations of the gauge group. In section 3 we give the general form of Ward identities for variations of adjoint and bifundamental fields.  In section 4 we study the simplest example of non-affine quiver theories, we write the anomalous Konishi reparameterizations and obtain a quadratic and a cubic constraints for the resolvents of this theory. 
In section 5 we derive the quadratic and cubic constraints for $A_n$ theories with  generic $n$, and sketch a procedure on how to calculate all higher-degree equations. We also evaluate, as a function of the resolvents, the expectation values of chiral operators with two and four bifundamental insertions in $\tr \frac{{\cal WW}}{z-\Phi}$.  In section 6 and 7 we study affine theories, starting again with the simplest example $\hat{A}_1$ and generalizing to generic $n$. In section 8 we give our conclusions and comment about open questions that might be interesting to answer.

\vspace{12pt}
\noindent{\it Note added}\\
\noindent While this work was being completed \cite{Naculich:2003cz} appeared on the net. One section of that paper has substantial overlaps with our example in section \ref{A2}. We obtain the same results for the quadratic and cubic constraints for the resolvents of $A_2$ non-affine quiver theories.

\subsection{Lightning review of quiver gauge theories}\label{review}
We are interested in gauge theories with gauge group\footnote{For simplicity we will only consider quivers associated to $A$-type Lie algebras.}
\begin{equation}\label{group}
U(N_1)\times U(N_2)\times U(N_3)\times\ldots\times U(N_n)
\end{equation}
and matter in the adjoint and bifundamental representations of these groups.

To the Lie algebra of (\ref{group}) is associated a simply-laced Dynkin diagram. We represent the theory by putting on each node a gauge group, and on each oriented link a bifundamental transforming in the fundamental of the gauge group the link starts from, and in the anti-fundamental of the one on which it goes to.

One choice for the superpotential for such theories is \cite{quivers}
\be\label{conf}
W_{tree}=\sum \tr \left( \bar{X}_{k+1,k}\Phi_{(k)}{X}_{k,k+1} - {X}_{k,k+1}\Phi_{(k+1)} \bar{X}_{k+1,k}\right)
\ee
where the sum runs from 1 to $n-1$ if there is no bifundamental linking the first and the $n^{th}$ node ($A_{n}$ quivers) or $n$ otherwise (affine $\hat{A}_{n-1}$ quivers).
 
The theory thus obtained has ${\cal N}=2$ supersymmetries.

${\cal N}=2$ supersymmetry can be broken down to ${\cal N}=1$ by perturbing $W_{tree}$ (\ref{conf}) with a polynomial term for the adjoint matter fields
\be\label{superpotential}
W_{tree}=\sum_{k=1}^{n}\tr W_{(k)}(\Phi_{(k)})+\sum_k \tr \left( \bar{X}_{k+1,k}\Phi_{(k)}{X}_{k,k+1} - {X}_{k,k+1}\Phi_{(k+1)} \bar{X}_{k+1,k}\right)
\ee
where again the extrema of the second sum depend on whether we are considering $A_{n}$ or $\hat{A}_{n-1}$, and
\be
W_{(k)}(\Phi_{(k)})=\sum_{p=1}^{n_k+1}\frac{g_{(k)p}}{p}\,\Phi_{(k)}^p
\ee

In the affine case, the superpotential needs to satisfy the condition \cite{CKV} \cite{tatar}
 \be\label{condWaff}
\sum_{k=1}^nW_{(k)}(z)=0
\ee
and indices are always considered mod $n$: $k+n\simeq k$.

One way of looking at affine quiver gauge theories in string theory, is as the low-energy description of the dynamics of D3-branes placed at points of $\mathbb{R}^4/\Gamma$ orbifolds \cite{quivers}.

D3-branes placed at points of these spaces can be described, at low energy, by the truncation of ${\cal N} =4$ SYM to the sector invariant under the orbifold action (for simplicity we take $\Gamma=\mathbb{Z}_n$)
\be
g(z^1,z^2)=(e^\frac{2\pi i}{n}z^1,e^{-\frac{2\pi i}{n}}z^2) 
\ee 
where $z^1$ and $z^2$ are the $\sigma$-model scalar fields with target $\mathbb{C}^2$. An action for their world-sheet fermionic partners can be written in an analogous way.

The group $\Gamma$ can be associated to the simply-laced Dynkin diagram associated with the simple Lie algebra of the gauge group, which, in the $\mathbb{Z}_n$ case, is $\hat{A}_{n-1}$.

To build non-affine quiver theories $A_{n-1}$ with this method we have to turn off one of the nodes, i.e. putting $N_n=0$.

Another way of building quiver gauge theories is by wrapping D5-branes around the two-cycles of ALE spaces. To treat orbifold singularities in $\mathbb{C}^2/{\mathbb{Z}_n}$ we can deform the space to an ALE space.

ALE spaces are 4-dimensional manifolds asymptotic to $\mathbb{R}^4/\Gamma$, where $\Gamma$ is a discrete subgroup of $SU(2)$. For the case $\Gamma=\mathbb{Z}_n$, the metric has been written explicitly \cite{Eguchi} as a function of $n$ parameters. 

$\mathbb{C}^2/{\mathbb{Z}_n}$ can be embedded in $\mathbb {C}^3$ as
\be
f=x^2+y^2+z^{n}=0
\ee

The origin is a singularity of this space since $f=df=0$, but the space can be desingularized. Let us introduce $n$ parameters $t_i$, and consider the spaces
\be\label{defor}
x^2+y^2+\prod_{i=1}^{r+1}(z+t_i)=0
\ee
with the constraint
\be
\sum_{i=1}^{r} t_i=0
\ee

In a sense, the deformation consists of blowing the singularity up into $n$ two-cycles, whose holomorphic volume is measured by the $t_i$'s, while keeping the asymptotic behavior unchanged. 

The roots of the Dynkin diagram associated with $A_n$ are identified with \cite{CKV}
\be
\alpha_i=t_i-t_{i+1}
\ee

From this point of view, $A_{n}$ quiver theories are the description of the low-energy dynamics of D5 branes wrapped over these two-cycles. Affine $\hat{A}_{n-1}$ quivers are obtained by adding D3-branes at points of the ALE space.

Quiver gauge theories have several different classical branches. Let us call $A_{(k)}\equiv \bar{X}_{k,k-1}X_{k-1,k}$ and $B_{(k)}\equiv -X_{k,k+1}\bar{X}_{k+1,k}$. The coulomb branch is characterized by 
\be\label{eigenC}
\begin{split}
&W^\prime_{(k)}(\Phi_{(k)})=0\\
&A_{(k)}=0\qquad B_{(k)}=0
\end{split}
\ee
for $ k=1,2\ldots,n$.

On the other hand, Higgs branches of $A_{n}$ are characterized by \cite{CKV} \cite{tatar} (for some $(k,j,l)$ with $k=1\ldots, n-2$; $j=1,\ldots,k$; $l=j+1,\ldots,n-1$)\footnote{The operators $X_j$, $Y_j$ and $\Phi_j$ commute classically because of the equations of motion, and can thus be diagonalized simultaneously. Lower case letters stand for the eigenvalues of such operators \cite{CKV} \cite{tatar}.}
\be\label{eigenH}
\begin{split}
&a_{(k)}=t_k-t_j\\
&b_{(k)}=t_l-t_{k+1}\\
&a_{(m)}=t_m\\
&b_{(m)}=-t_{m+1}\\
&\sum_{m=j}^{l-1}W^\prime_{(m)}(\phi)=0
\end{split}
\ee
where the $t_i$'s are as defined in (\ref{defor}), $\phi$ is the eigenvalue of $\Phi_{(m)}$, and $m=j,\ldots,l-1$.

For each gauge group $U(N_k)$ we can choose the operators $\Phi_{(k)}$, $A_{(k)}$ and $B_{(k)}$ to have $s_1$ eigenvalues $(\phi_{(k,1)},a_{(k,1)},b_{(k,1)})$, $s_2$ eigenvalues $(\phi_{(k,2)},a_{(k,2)},b_{(k,2)})$ and so on, picking from the allowed eigenvalue combinations (\ref{eigenC}) or (\ref{eigenH}). The total gauge group is then semiclassically broken:

\be
\prod_{k=1}^{n-1}U(N_k)\quad\rightarrow\quad \prod_{i} U(M_i)
\ee

A similar analysis can be carried out for affine theories. What is found is again that the gauge group is broken, this time to
\be
\prod_{k=1}^{n}U(N_k)\quad\rightarrow\quad U(N)\times\prod_{i} U(M_i)\quad\rightarrow\quad U(1)^N\times\prod_{i} U(M_i)
\ee
where $N$ is the number of imaginary positive roots of the affine A-D-E \cite{CKV} \cite{tatar}, and the second arrow represents the higgsing of $U(N)$ for a generic point on the moduli space of the $U(N)$ theory.

Quantum effects dynamically break each of these classically surviving symmetries in a free $U(1)$ times a strongly coupled, confining $SU(M_{i})$.

The low energy effective degrees of freedom are given by
\be\label{oper1}
\begin{split}
&N_{k,i}=\tr_{U(N_{k,i})}\mathds{1}\\
&w^\alpha_{k,i}=\frac{1}{4\pi}\tr_{U(N_{k,i})}{\cal W}^{(k,i)\alpha}\\
&S_{k,i}=-\fr\tr_{U(N_{k,i})}{{\cal W}_{\alpha}^{(k,i)}{\cal W}^{(k,i)\alpha}}
\end{split}
\ee
There is a much more compact and efficient way of writing these degrees of freedom which makes use of holomorphicity \cite{CDSW}. For each gauge group $U(N_k)$ let us introduce the complex functions
\be
\begin{split}
&T_{(k)}(z)\equiv \tr \frac{1}{z-\Phi_{(k)}} \\
&\rho^\alpha_{(k)}(z)\equiv \frac{1}{4\pi}\tr\frac{{\cal W}^{(k)\alpha}}{z-\Phi_{(k)}}\\
&R_{(k)}(z)\equiv -\fr\tr\k
\end{split}
\ee
The operators in (\ref{oper1}) can then be written as
\be
\begin{split}
&N_{k,i}= -\frac{1}{2\pi i}\oint_{C_{k,i}}dz\,T_{(k)}(z)\\
&w^\alpha_{k,i}=-\frac{1}{2\pi i}\oint_{C_{k,i}}dz\,\rho^\alpha_{(k)}(z)\\
&S_{k,i}= -\frac{1}{2\pi i}\oint_{C_{k,i}}dz\,R_{(k)}(z)
\end{split}
\ee 
where $C_{k,i}$ is any closed path that encircles the $i^{th}$ critical point of $W_k(z)$ and no other critical point.


\newsection{Chiral ring rules} 

Chiral operators\footnote{The part of this section concerning general relations in the chiral ring and relations involving adjoint operators,  follows closely \cite{CDSW}.}  are annihilated by the supersymmetry charges $\bar{{\cal Q}}_{\dot{\alpha}}$ of one chirality
\be
\{ \bar{{\cal Q}}_{\dot{\alpha}}, {\cal O}\} = 0
\ee

An equivalence relation can be introduced in the set of chiral operators: two operators are equivalent if they differ by a \q--commutator $\{\q,...\}$. Since the vacuum is also annihilated by the supersymmetry generators, two operators which are equivalent under this relation, will have the same expectation value on the vacuum. Since the equivalence class of the product of two chiral operators (which is itself a chiral operator) is equal to the product of the equivalence classes of the single operators, we can introduce a ring structure on the set of chiral operators, which has been conveniently called chiral ring.

Some useful relations hold in the chiral ring. First of all the expectation value of a product of chiral operators is independent of the position of each operator
\be
\frac{\partial}{\partial x_j^\mu} \langle {\cal O}_{i_1}(x_1)\ldots {\cal O}_{i_n}(x_n) \rangle=0 \qquad \mathrm{for~ all~}j=1,2,\ldots n
\ee

This fact, associated to cluster decomposition, implies that the correlation functions of chiral operators factorize
\be
\langle {\cal O}_{i_1}(x_1)\ldots {\cal O}_{i_n}(x_n) \rangle=\langle {\cal O}_{i_1}\rangle \ldots \langle {\cal O}_{i_n} \rangle
\ee
where there is no need to specify the position where operators are inserted.

Some more relations hold among chiral operators, which we need to consider not to overcount elements of the chiral ring. Let us consider a theory with two gauge groups $U(N_1)$ and $U(N_2)$, let $\Phi_{(1)}$ (respectively $\Phi_{(2)}$) be a field in the adjoint of the first (second) gauge group, and $X_{12}$ ($\bar{X}_{21}$) be in the fundamental (antifundamental) representation of the first gauge group and in the antifundamental (fundamental) of the second group. Using the definition of the field strength \W ~of the gauge groups as a superfield
\be
\W=\{\q,D_{\alpha\dot{\alpha}}\}
\ee
its action on an adjoint operator of one of the two groups is given by
\be \label{fieldstr}
[ \W,{\cal O}\}=[\q ,D_{\alpha\dot{\alpha}}^{(i)}{\cal O}\}\c= 0
\ee
where \c= indicates that the equality is only valid in the chiral ring.

The action of the field strength on the bifundamentals is a little more delicate. Again considering ${\cal W}_\alpha $ as a superfield we may write
\be\label{fieldstrbif}
\W X_{12}=\{\q ,D_{\alpha\dot{\alpha}}^{(i)}X_{12}\}\c= 0   
\ee

We can also represent $X_{12}$ as the tensor product of a fundamental element of the first gauge group and an antifundamental element of the second group
\be
X_{12}=v\otimes\bar{u}
\ee
and write the action of the field strength as
\be\label{crbif}
\begin{split}
\W B &  = \W^{(1)a_1} \rho(T_{(1)}^{a_1})(v\otimes\bar{u})+\W^{(2)a_2} \rho(T_{(2)}^{a_2})(v\otimes\bar{u})= \\
& = \W^{(1)a_1} (T_{(1)}^{a_1}v)\otimes\bar{u}+\W^{(2)a_2} v \otimes(\bar{u}T_{(2)}^{a_2})=\\
& =(\W^{(1)}v)\otimes \bar{u}+v\otimes(\bar{u}\W^{(2)})=\\
& =\W^{(1)} X_{12}+X_{12}\W^{(2)}\c= 0
\end{split}
\ee
where last equation follows from (\ref{fieldstrbif}) and $\rho(T)$ indicates the realization of the generator of the algebra in the representation considered.

It is straightforward to show that these relations can be extended to the case in which we have a theory with $n$ gauge groups $U(N_1),\; U(N_2),\ldots U(N_n)$ with matter in the adjoint or bifundamental representation of two adjacent groups.

If in (\ref{fieldstr}) we take ${\cal O}$ to be the field strength of one of the groups we find
\be
\{\W^{(i)},{\cal W}_\beta^{(i)}\}\c= 0
\ee
where obviously the two fields are in the adjoint of the same group. Since $\alpha=1,2$ this relation implies that any string of ${\cal W}$'s longer than 2 is equal to zero in the chiral ring \cite{CDSW}. The operators we are interested in will contain, then, up to two ${\cal W}$ insertions.

If, instead, we substitute ${\cal O}=\Phi_{(i)}$ then
\be \label{cradj}
[\W^{(i)},\Phi_{(i)}]\c= 0
\ee

The chiral ring of this quiver theory is then composed of the following operators
\be
\begin{split}
&\tr \left( \Phi_{(i_1)}^{p_1} Y^{q_1}\ldots\Phi_{(i_k)}^{p_k} Y^{q_k}\right)\\
&\tr \left(\W^{(i_1)}\Phi_{(i_1)}^{p_1} Y^{q_1}\ldots\Phi_{(i_k)}^{p_k} Y^{q_k}\right)\\
&\tr  \left(\W^{(i_1)} {\cal W}^{(i_1)\alpha} \Phi_{(i_1)}^{p_1} Y^{q_1}\ldots\Phi_{(i_k)}^{p_k} Y^{q_k}\right)
\end{split}
\ee
where not to make the formulas too clumsy, we have omitted indices on the bifundamental fields and indicated with $Y$ both $X$ and $\bar{X}$ fields. Their index structure is fixed, in any case, by the gauge invariance of the operators, and by the rule that $X$ fields always go from a lower group to a higher one, while $\bar{X}$ goes the other way. Because of (\ref{crbif}) and (\ref{cradj}) the position of ${\cal W}$ insertions does not really matter. On the contrary $\Phi$'s and bifundamentals do not commute, making their position relevant to the classification of operators.


\newsection{Generalized Konishi anomaly with bifundamental matter}

In the presence of matter in the bifundamental representation, the generalized Konishi anomaly equation is similar to the equations obtained in \cite{CDSW} and \cite{Seiberg} for adjoint and fundamental matter.

Let
\be
\begin{split}
& \delta \Phi =f_\Phi (\Phi,\W,X,\bar{X})\\
& \delta X=f_X (\Phi,\W,X,\bar{X})\\
& \delta \bar{X}=f_{\bar{X}} (\Phi,\W,X,\bar{X})
\end{split}
\ee
be the variations of the matter fields, then the anomaly equations read
\be\label{konishi}
\begin{split}
&\left\langle \tr f_\Phi(\Phi,\W,X,\bar{X})\frac{\partial W_{tree}}{\partial \Phi}\right\rangle=-\frac{1}{32\pi^2}\left\langle\sum_{i,j}\left( \left[ \W,\left[{\cal W}^{\alpha},\frac{\partial f_\Phi (\Phi,\W,X,\bar{X})}{\partial \Phi_{ij}}  \right]   \right]  \right)_{ij}\right\rangle\\
&\left\langle \tr f_X(\Phi,\W,X,\bar{X})\frac{\partial W_{tree}}{\partial X}\right\rangle=-\frac{1}{32\pi^2}\left\langle\sum_{i,j}\left( \W{\cal W}^{\alpha}\frac{\partial f_X (\Phi,\W,X,\bar{X})}{\partial X_{ij}}  \right)_{ij}\right\rangle\\
&\left\langle \tr f_{\bar{X}}(\Phi,\W,X,\bar{X})\frac{\partial W_{tree}}{\partial \bar{X}}\right\rangle=-\frac{1}{32\pi^2}\left\langle\sum_{i,j}\left( \W{\cal W}^{\alpha}\frac{\partial f_{\bar{X}} (\Phi,\W,X,\bar{X})}{\partial \bar{X}_{ij}}  \right)_{ij}\right\rangle\\
\end{split}
\ee
where we have suppressed gauge group indices, but again they are fixed by gauge invariance once we choose which field we are deforming.

An identity we will find really useful in the evaluation of the anomaly equation coming from the variation of adjoint matter is the following\footnote{The demonstration of this relation is quite similar to the proof of the identity in \cite{CDSW} which is the same as (\ref{identity}) with $\chi_3=1$.}
\be\label{identity}
\sum_{i,j}\left[\chi_1,\left[\chi_2,\frac{\partial}{\partial \Phi_{ij}} \frac{\W{\cal W}^{\alpha}}{z-\Phi} \chi_3\right]\right]_{ij}=\left( \tr \frac{\chi_1\chi_2}{z-\Phi}\right)\left( \tr \frac{\chi_1\chi_2}{z-\Phi}\chi_3\right)
\ee
where $\chi_1^2=\chi_2^2=0$, $\Phi$ commutes with $\chi_1$ and $\chi_2$ but need not commute with $\chi_3$.

Another important property of the operators we will consider, which we will often use, is a generalized form of the cyclical property of traces. Let us consider the combinations $X_{k,k+1}\bar{X}_{k+1,k}$ and $\bar{X}_{k+1,k}X_{k,k+1}$, they are operators in the adjoint of $U(N_k)$ and $U(N_{k+1})$ respectively, thus when we write $\tr X_{k,k+1}\bar{X}_{k+1,k}$ and $\tr\bar{X}_{k+1,k}X_{k,k+1}$ it is meant that we are taking the trace on two different groups. Nevertheless these operators are identical. In fact let us write them in components ($I,\bar{I}$ will be indices in the fundamental and antifundamental of $U(N_k)$ respectively, while $j,\bar{j}$ will be in the fundamental and antifundamental of $U(N_{k+1})$)
\be\label{cicl}
\tr_{U(N_k)} X \bar{X}= X_{I\bar{j}}\bar{X}_{j\bar{I}}=\bar{X}_{j\bar{I}}X_{I\bar{j}}=\tr_{U(N_{k+1})}\bar{X}X
\ee
where gauge group indices have been neglected and sums over fundamental - antifundamental pairs of indices are understood. This generalized cyclical property of traces is still valid when adjoint fields are inserted.


\newsection{$A_2$ quiver}\label{A2}
We start by considering the $A_2$ quiver gauge theory with gauge group $U(N_1) 
\times U(N_2)$. As already said in section \ref{review}, the chiral matter content of this theory is made up of two fields $\Phi_1$ and $\Phi_2$ and a couple of bifundamental fields $X$ and $\bar{X}$.        

For convenience, we rewrite here (\ref{superpotential}) for the $A_2$ case
\be
W_{tree}= \tr W_{(1)}(\Phi_{(1)}) + \tr W_{(2)}(\Phi_{(2)}) + \tr \bar{X} \Phi_{(1)} X -
\tr X \Phi_{(2)} \bar{X}
\ee

Following \cite{CDSW} and \cite{Seiberg} we consider a set of independent 
transformations 
for the chiral fields and then write down the anomaly equations for these 
transformations. 

Combining them properly we can 
obtain two equations, one quadratic and one cubic, which depend only on 
the resolvents $R_{(1)}$ and $R_{(2)}$. This result can be generalized to an 
arbitrary quiver theory $A_{n}$ as we discuss in section \ref{general}.

\vspace{12pt}
{\noindent {\it Quadratic equation}}
\vspace{12pt}

To obtain the quadratic equation, we need the three transformations
\begin{align}
\delta \Phi_{(1)}=& \frac{{\cal W}_\alpha^{(1)} 
{\cal W}^{(1)\alpha}}
{z-\Phi_{(1)}} \nonumber \\
\delta \Phi_{(2)}=& \frac{{\cal W}_\alpha^{(2)}{\cal W}^{(2)\alpha}}{z-\Phi_{(2)}} \\
\delta X =&  \frac{{\cal W}_\alpha^{(1)}{\cal W}^{(1)\alpha}}
{z-\Phi_{(1)}} X \frac{1}{z-\Phi_{(2)}} \nonumber 
\end{align}

The corresponding anomaly equations are
\begin{align}
\label{eq:anomalia1}
- \frac{1}{32\pi^2} \tr\left( W'_{(1)}(\Phi_{(1)})\frac{{\cal W}_\alpha^{(1)} 
{\cal W}^{{(1)} \alpha}}{z-\Phi_{(1)}} + \bar{X} \frac{{\cal W}_\alpha^{(1)} 
{\cal W}^{{(1)} \alpha}}{z-\Phi_{(1)}} X \right) =& R_{(1)}^2(z) \nonumber \\  
  - \frac{1}{32\pi^2} \tr\left( W'_{(2)}(\Phi_{(2)})\frac{{\cal W}_\alpha^{(2)}{\cal W}^{{(2)} \alpha}}{z-\Phi_{(2)}} - X \frac{{\cal W}_\alpha^{(2)} 
{\cal W}^{{(2)} \alpha}}{z-\Phi_{(2)}} \bar{X} \right) =& R_{(2)}^2(z) \\
- \frac{1}{32\pi^2} \tr
\left( \bar{X} \frac{{\cal W}_\alpha^{(1)} 
{\cal W}^{{(1)} \alpha}}{z-\Phi_{(1)}} X - X \frac{{\cal W}_\alpha^{(2)} 
{\cal W}^{{(2)} \alpha}}{z-\Phi_{(2)}} \bar{X} \right) =&R_{(1)}(z) R_{(2)}(z). \nonumber
\end{align}
We can rewrite the first two equations defining, as usual \cite{CDSW} \cite{Seiberg}, the two polynomials ($k=1,2$) 
\be
\label{deff}
f_{(k)}(z) = -\frac{1}{8\pi^2} \tr \frac{W'_{(k)}(\Phi_k)-W'_{(k)}(z)}{z-\Phi_{(k)}}
{\cal W}_\alpha^{(k)}{\cal W}^{{(k)} \alpha}.
\ee
All terms containing two fields in the bifundamental 
cancel if we 
take the sum of the first two equations in (\ref{eq:anomalia1}) minus the 
third one. As a result we obtain
\be
\label{quadraticaa2}
W'_{(1)}(z)R_{(1)}(z)+\frac{1}{4}f_{(1)}(z) + W'_{(2)}(z)R_{(2)}(z)+\frac{1}{4}f_{(2)}(z) + R_{(1)}(z) 
R_{(2)}(z) = R_{(1)}(z)^2 + R_{(2)}(z)^2
\ee
which is quadratic in $R_{(k)}(z)$. 

From (\ref{eq:anomalia1}) we can also calculate the expectation values of double $X$ operators
\be\label{A2XX}
\begin{split}
&-\fr\tr X \frac{{\cal W}_\alpha^{(2)}{\cal W}^{(2)}_\alpha}{z-\Phi_{(2)}}\bar{X}=R_{(1)}^2-W^\prime_{(1)}R_{(1)}-\frac{1}{4}f_{(1)}-R_{(1)}R_{(2)}\\
&-\fr\tr \bar{X} \frac{{\cal W}_\alpha^{(2)}{\cal W}^{(2)}_\alpha}{z-\Phi_{(1)}}X=-(R_{(2)}^2-W^\prime_{(2)}R_{(2)}-\frac{1}{4}f_{(2)}-R_{(1)}R_{(2)})
\end{split}
\ee

\vspace{12pt}
{\noindent {\it Cubic equation}}
\vspace{12pt}

The cubic equation can be derived in a similar way, considering a little more
 complicated transformations 
\begin{align}
\delta \Phi_{(1)}=&   \frac{{\cal W}_\alpha {\cal W}^\alpha}{z-\Phi_{(1)}} 
X \bar{X} \nonumber \\
\delta \Phi_{(2)}=& \frac{{\cal W}_\alpha{\cal W}^\alpha}
{z-\Phi_{(2)}} \bar{X} X  \\
\delta X=& \frac{{\cal W}_\alpha{\cal W}^\alpha}
{z-\Phi_{(1)}} X \bar{X} X \frac{1}{z-\Phi_{(2)}} \nonumber
\end{align}

Again we write the corresponding anomaly equations 
\begin{align}
\label{eq:anomalia2}
- \frac{1}{32\pi^2} \tr& \left( W'_{(1)}(\Phi_{(1)})
\frac{\ensuremath{{\cal W}_\alpha^{(1)}} 
\ensuremath{{\cal W}^{{(1)} \alpha}}}{z-\Phi_{(1)}}X\bar{X} 
+  X\bar{X} 
\frac{\ensuremath{{\cal W}_\alpha^{(1)}}
\ensuremath{{\cal W}^{{(1)} \alpha}}}{z-\Phi_{(1)}} X\bar{X} \right) 
=\nonumber \\
\qquad =-&\frac{1}{32\pi^2} R_{(1)}\tr \bar{X}\frac{{\cal W}_\alpha^{(1)} 
{\cal W}^{{(1)}\alpha}}{z-\Phi_{(1)}} X  \nonumber 
\end{align}
\begin{align}
- \frac{1}{32\pi^2} \tr& \left( W'_{(2)}(\Phi_{(2)})
\frac{{\cal W}_\alpha^{(2)}{\cal W}^{{(2)} \alpha}}{z-\Phi_{(2)}} \bar{X} X 
-\bar{X} X 
\frac{{\cal W}_\alpha^{(2)}{\cal W}^{{(2)} \alpha}}{z-\Phi_{(2)}}
\bar{X}X \right) = \nonumber \\
\qquad= -&\frac{1}{32\pi^2} R_{(2)}(z) \tr  X \frac{{\cal W}_\alpha^{(2)} 
{\cal W}^{{(2)}\alpha}} {z-\Phi_{(2)}} \bar{X} 
\end{align}
\begin{align}
- \frac{1}{32\pi^2} \tr& \left( X \bar{X} 
\frac{{\cal W}_\alpha^{(1)}{\cal W}^{{(1)} \alpha}}{z-\Phi_{(1)}} X \bar{X}
- \bar{X} X \frac{{\cal W}_\alpha^{(2)}{\cal W}^{{(2)} \alpha}}{z-\Phi_{(2)}}
\bar{X} X \right) = \nonumber \\ 
\qquad =-&\frac{1}{32\pi^2} \left( R_{(1)}(z)\tr X \frac{{\cal W}_\alpha^{(2)} 
{\cal W}^{{(2)}\alpha}} {z-\Phi_{(2)}} \bar{X} + R_{(2)}(z) \tr\bar{X}
\frac{{\cal W}_\alpha^{(1)}{\cal W}^{{(1)}\alpha}}{z-\Phi_{(1)}} X \right).
\nonumber
\end{align}

As in (\ref{deff}), we define two polynomials
\begin{align}
g_{(1)}(z) &= -\frac{1}{8\pi^2} \tr \left( \frac{W'_{(1)}(\Phi_{(1)})-W'_{(1)}
(z)}{z-\Phi_{(1)}}
{\cal W}_\alpha^{(1)}{\cal W}^{{(1)}\alpha} X\bar{X}\right) \nonumber\\
g_{(2)}(z) &= -\frac{1}{8\pi^2} \tr \left( \frac{W'_{(2)}(\Phi_{(2)})-W'_{(2)}
(z)}{z-\Phi_{(2)}}
{\cal W}_\alpha^{(2)}{\cal W}^{{(2)}\alpha} \bar{X}X\right) 
\end{align} 
to eliminate the terms $W'_k(\Phi_k)$ from the first two Ward identities 
(\ref{eq:anomalia2}), and make use of (\ref{A2XX}) to rewrite the operators
 that depend 
only on two bifundamentals in terms of the resolvents 
$R_{(k)}(z)$. Finally we cancel the operators with four 
bifundamentals subtracting the third equation of (\ref{eq:anomalia2}) from the 
sum of the first two. The result is the cubic equation
\begin{align}
\label{cubicaa2}
&R_{(1)}(z)^2R_{(2)}(z)- R_{(1)}(z)R_{(2)}(z)^2 = W'_{(1)}(z) 
\left( R_{(1)}(z)^2 - W'_{(1)}(z) R_{(1)}(z) 
-\frac{1}{4} f_{(1)}(z) \right)+ \nonumber \\
&-  W'_{(2)}(z) \left( R_{(2)}(z)^2 - W'_{(2)}(z) R_{(2)}(z) 
-\frac{1}{4} f_{(2)}(z) \right) + \frac{1}{4} g_{(1)}(z)+
\frac{1}{4} g_{(2)}(z).
\end{align} 

\vspace{12pt}
{\noindent {\it Spectral curve}}
\vspace{12pt}

Following \cite{Dijkgraaf:2002vw} and \cite{Lazaroiu} we may use (\ref{quadraticaa2}) and (\ref{cubicaa2}) to write a cubic equation whose roots are related to the resolvents $R_{(1)}$ and $R_{(2)}$. In the limit of turning off the tree potential for the adjoint fields, in which we should recover ${\cal N}=2$ supersymmetry \cite{Cach}, after going on-shell (e.g. minimizing the effective superpotential), this equation should be related with the Seiberg-Witten curve for $A_2$ quiver theories.


We can eliminate terms linear in $R_{(k)}$ from the quadratic equation (\ref{quadraticaa2}) making 
the following change of variables: 
\begin{align}
a_{(1)}(z)&= R_{(1)}(z) - \frac{2}{3} W'_{(1)}(z) - \frac{1}{3} W'_{(2)}(z) \nonumber \\
a_{(2)}(z)&= -R_{(2)}(z) + \frac{1}{3} W'_{(1)}(z) + \frac{2}{3} W'_{(2)}(z).
\end{align}

With this choice the two equations are:
\begin{align}\label{modeq}
a_{(1)}(z)^2 + a_{(2)}(z)^2 +a_{(1)}(z)a_{(2)}(z) - p^{(2)}(z) = 0 \nonumber \\
a_{(1)}(z)^2 a_{(2)}(z) + a_{(1)}(z) a_{(2)}(z)^2 + p^{(3)}(z) = 0
\end{align}
where
\begin{align}
p^{(2)}(z) \equiv \,&t_{(1)}(z)^2 -  t_{(1)}(z)t_{(2)}(z)+t_{(2)}(z)^2 +\frac{1}{4} f_{(1)}(z) +\frac{1}{4}f_{(2)}(z) 
\nonumber \\
p^{(3)}(z) \equiv &-t_{(1)}(z)t_{(2)}(z)( t_{(1)}(z)- t_{(2)}(z) ) +\frac{1}{4} t_{(1)}(z)f_{(2)}(z) 
-\frac{1}{4} t_{(2)}(z)f_{(1)}(z)+\\\nonumber&+
\frac{1}{4}g_{(1)}(z) +\frac{1}{4} g_{(2)}(z)
\end{align}
and
\begin{align}
t_{(1)}(z) &\equiv \frac{2}{3} W'_{(1)}(z) + \frac{1}{3} W'_{(2)}(z) \nonumber \\
t_{(2)}(z) &\equiv \frac{1}{3} W'_{(1)}(z) + \frac{2}{3} W'_{(2)}(z).
\end{align}
If we define 
\be
a_{(3)} \equiv -a_{(1)} - a_{(2)}
\end{equation}
equations (\ref{modeq}) can be rewritten as
\begin{align}
&\sum_{i<j}^3 a_{(i)} a_{(j)}=p^{(2)} \nonumber \\
&\sum_{i<j<k}^3a_{(i)}a_{(j)}a_{(k)} =p^{(3)}  
\end{align}
which are the conditions satisfied by the three roots $a_{(i)}$ of a cubic equation of the form
\be
y^3- p^{(2)}y - p^{(3)}=0.
\end{equation}
This is exactly the curve suggested in \cite{Dijkgraaf:2002vw} and calculated from the matrix model in \cite{Lazaroiu}.


\newsection{Generalization to $A_{n}$}\label{general}
An interesting problem is to generalize the procedure we followed in the simple case $A_2$
to the more involved $A_n$ quiver theory, with $n$ generic. Even though the problem seems
too complicated to make it possible to give an answer for the general case, we will write the exact quadratic and cubic 
equations for the resolvents, obtain the expectation values of operators with two insertions 
of field strength and two or four bifundamental fields, and sketch a general procedure to obtain all
the equations for a fixed value of $n$.

\vspace{12pt}
{\it Quadratic equation}
\vspace{12pt}

We consider the following transformations for the adjoint matter fields
\be\label{adj}
\delta\Phi_{(k)}=\frac{{\cal W}_{\alpha}^{(k)}{\cal W}^{(k)\alpha}}{z-\Phi_{(k)}}
\ee
These variations are the analogue of the variations considered in \cite{CDSW}, for the case of the direct product of different gauge groups.
From (\ref{konishi}) we obtain the following equations $(k=2,\ldots n-1)$
\begin{align}
&{W}^\prime_{(k)}(z)R_{(k)}(z)+\frac{1}{4}f_{(k)}(z)-\frac{1}{32\pi^2}\left( \tr\bar{X} \k X-\tr X \k\bar{X}\right)  =\label{anadjk}\\&\qquad\qquad=R_{(k)}(z)^2\nonumber\\
&{W}^\prime_{(1)}(z)R_{(1)}(z)+\frac{1}{4}f_{(1)}(z)-\frac{1}{32\pi^2}\tr\bar{X}\p1 X = R_{(1)}(z)^2\label{anadj1}\\
&{W}^\prime_{(n)}(z)R_{(n)}(z)+\frac{1}{4}f_{(n)}(z)+\frac{1}{32\pi^2}\tr X\pn\bar{X} =R_{(n)}(z)^2 \label{anadjn}
\end{align}
where, as for $A_2$, we have defined the holomorphic polynomial functions of degree $n_{k}-1$
\be\label{fz}
f_{(k)}(z)\equiv -\frac{1}{8\pi^2}\tr\left(  \frac{ W^\prime_{(k)}(\Phi_{(k)})-W^\prime_{(k)}(z)}{z-\Phi_{(k)}}{\cal W}_{\alpha}^{(k)}{\cal W}^{(k)\alpha}\right)
\ee
Taking the sum of all of the above equations (\ref{anadjk})-(\ref{anadjn}) we find
\be\label{quad1}
\sum_{k=1}^n\left(\!W^\prime_{(k)}R_{(k)}\!+\!\frac{1}{4}f_{(k)} \right) -\frac{1}{32\pi^2}\sum_{k=1}^{n-1}\tr \!\left(\bar{X}\k X\!-\!X\pk1\bar{X}   \right)=\sum_{k=1}^{n}R_{(k)}^2
\ee
where the $z$-dependence of the holomorphic functions $W^\prime_{(k)}(z)$, $R_{(k)}(z)$ and $f_{(k)}(z)$ is understood.

The $X$-$\bar{X}$ and $\bar{X}$-$X$ traces may be evaluated using the anomaly generated by
\be\label{bifund}
\delta X_{k,k+1}=\frac{{\cal W}_{\alpha}^{(k)}{\cal W}^{(k)\alpha}}{z-\Phi_{(k)}}X_{k,k+1}\frac{1}{z-\Phi_{(k+1)}}
\ee

Equation (\ref{bifund}) is peculiar of bifundamental fields: since we want to perturb 
the link between the $k^{th}$ and $(k+1)^{th}$ nodes of the quiver, we deform the 
two extrema of the link using adjoint fields\footnote{Note that because of the chiral
 ring relations (\ref{crbif}) and (\ref{cradj}), we would have obtained the same transformation 
 by moving one or both field strengths on the right of $X_{k,k+1}$.} and glue them with 
 bifundamentals.

Using the second of (\ref{konishi}) and (\ref{identity}) we find
\be\label{anbif}
-\fr\tr\bar{X}\k X+\fr\tr X\pk1\bar{X}=R_{(k)}R_{(k+1)}
\ee

substituting into (\ref{quad1}) we finally get
\be
\sum_{k=1}^n\left(W^\prime_{(k)}(z)R_{(k)}(z)+\frac{1}{4}f_{(k)}(z)\right)+\sum_{k=1}^{n-1}R_{(k)}(z)R_{(k+1)}(z)=\sum_{k=1}^n R_{(k)}(z)^2
\ee
which is the quadratic equation for the resolvents of a generic $A_n$ quiver theory.

The equations we have written so far can also be used to evaluate the expectation values of the $X$-$\bar{X}$ and $\bar{X}$-$X$ trace operators. In fact we can use (\ref{anadj1}) to obtain
\be
-\fr\tr\bar{X}\p1 X =R_{(1)}^2(z)-W^\prime_{(1)}(z)R_{(1)}(z)-\frac{1}{4}f_{(1)}(z)
\ee

Plugging this result into (\ref{anbif}) with $k=1$ we can then obtain $\tr X \frac{{\cal WW}}{z-\Phi_{(2)}}\bar{X}$, which can in turn be used into (\ref{anadjk}) for $k=2$, and so on. The final result is
\be
\begin{split}
-&\fr\tr\bar{X}\k X=\sum_{p=1}^{k}\left(R_{(p)}^2-W^\prime_{(p)}R_{(p)}-\frac{1}{4}f_{(p)}- R_{(p-1)}R_{(p)}\right)\\
-&\fr\tr X\pk1 \bar{X}=\sum_{p=1}^{k}\left(R_{(p)}^2-W^\prime_{(p)}R_{(p)}-\frac{1}{4}f_{(p)}- R_{(p)}R_{(p+1)}\right)
\end{split}
\ee

\vspace{12pt}
{\noindent\it Cubic equation}
\vspace{12pt}

As for the $A_2$ theory, to obtain an equation which is cubic in the resolvents, 
we need to consider more complicated transformations for $\phi$, $X$ and $\bar{X}$. 
We thus add a $X$-$\bar{X}$ couple to the transformations (\ref{adj}) and (\ref{bifund}). 
There is more than one way to do so, because $\Phi$, $X$ and $\bar{X}$ do not commute: for 
the $\Phi$ variation we could put one of the two impurities on the left of the adjoint block 
and one on the right, or both on either side, while for the $X$ variation we could use any of 
the combinations of two $X$'s and one $\bar{X}$ between the two adjoint blocks. We will first 
rule out some of these possibilities, and then keep all of the remaining and calculate the anomaly 
equations for all of them: they will be just enough to obtain a cubic equation and evaluate the 
expectation values of all 4-$X$ operators.

First of all, we will not consider the adjoint transformations having one link\footnote{By link we mean an $X$ or an $\bar{X}$ insertion either in the reparameterizations of the adjoint and bifundamental fields, or in the trace operators.} on 
the left and one on the right. This is obvious since to build such a variation the adjoint 
field we insert must be different from the adjoint field we are varying 
(e.g. $\delta\Phi_k=X\frac{{\cal WW}}{z-\Phi_{k+1}}\bar{X}$), thus the quantum contribution 
to the anomaly (the commutator part in the first of (\ref{konishi})) vanishes, and only the 
classical part contributes, which is telling us nothing more about the low energy dynamics of 
the theory, than the classical equations of motion do. The choices with both $X$ and $\bar{X}$ on 
the left, or both of them on the right are actually equivalent because of the cyclicity of the trace (\ref{cicl}). 
Thus we are left with only two possible variations for the adjoint fields, which correspond to the two 
orderings of the pair $(X,\bar{X})$ on the right of the adjoint block.

The variations we consider are then
\begin{subequations}\label{cubicvar}
\be
\label{a} \delta\Phi_{(k)}=\k X\bar{X}
\ee\be
\label{b} \delta\Phi_{(k)}=\k \bar{X}X
\ee\be
\label{c} \delta X_{k,k+1}=\k X\bar{X}X\frac{1}{z-\Phi_{(k+1)}}
\ee\be
\label{d} \delta X_{k,k+1}=\k\bar{X}XX\frac{1}{z-\Phi_{(k+1)}}
\ee\be
\label{e} \delta X_{k,k+1}=\k XX\bar{X}\frac{1}{z-\Phi_{(k+1)}}
\ee\be
\label{f} \delta \bar{X}_{k+1,k}=\pk1 \bar{X}X\bar{X}\frac{1}{z-\Phi_{(k)}}
\ee\be
\label{g} \delta \bar{X}_{k+1,k}=\pk1 \bar{X}\bar{X}X\frac{1}{z-\Phi_{(k)}}
\ee\be
\label{h} \delta \bar{X}_{k+1,k}=\pk1 X\bar{X}\bar{X}\frac{1}{z-\Phi_{(k)}}
\ee
\end{subequations}

The Ward identities that follow from them are
\begin{subequations}\label{cubicward}
\be\begin{split}
{\scriptstyle{(\ref{a}),\: k=1}} \quad\rightarrow\quad & (W^\prime_{(1)}-R_{(1)})(R_{(1)}^2-W^\prime_{(1)}R_{(1)}-\frac{1}{4}f_{(1)})+\frac{1}{4}f_{(1)}^{1,1}=\\&=\fr\tr X\bar{X}\p1 X\bar{X}\label{wa}
\end{split}\ee
\be\begin{split}
{\scriptstyle{k=2,\ldots n-1 }}\quad\rightarrow\quad & (W^\prime_{(k)}-R_{(k)})\sum_{p=1}^k(R_{(p)}^2-W^\prime_{(p)}R_{(p)}-\frac{1}{4}f_{(p)}-R_{(p-1)}R_{(p)})+\frac{1}{4}f_{(k)}^{1,1}=\\&=\fr\tr \left(X\bar{X}\k X\bar{X} - \bar{X}X\k X\bar{X}\right)\label{wb}
\end{split}\ee
\be\begin{split}
{\scriptstyle{(\ref{b}),\: k=n }}\quad\rightarrow\quad & (W^\prime_{(n)}-R_{(n)})\sum_{p=1}^{n-1}(R_{(p)}^2-W^\prime_{(p)}R_{(p)}-\frac{1}{4}f_{(p)}-R_{(p)}R_{(p+1)})+\frac{1}{4}f_{(n)}^{1,2}=\\&=-\fr\tr \bar{X}X\pn \bar{X}X\label{wc}
\end{split}\ee
\be\begin{split}
{\scriptstyle{k=2,\ldots n-1}}\quad \rightarrow\quad & (W^\prime_{(k)}-R_{(k)})\sum_{p=1}^{k-1}(R_{(p)}^2-W^\prime_{(p)}R_{(p)}-\frac{1}{4}f_{(p)}-R_{(p)}R_{(p+1)})+\frac{1}{4}f_{(k)}^{1,2}=\\&=\fr\tr \left(X\bar{X}\k \bar{X}X -\bar{X}X\k \bar{X}X\right)\label{wd}
\end{split}\ee
\be\begin{split}
{\scriptstyle{(\ref{c})}}\quad \rightarrow\quad & R_{(k+1)}\sum_{p=1}^k(R_{(p)}^2-W^\prime_{(p)}R_{(p)}-\frac{1}{4}f_{(p)}-R_{(p-1)}R_{(p)})+\\&+R_{(k)}\sum_{p=1}^k(R_{(p)}^2-W^\prime_{(p)}R_{(p)}-\frac{1}{4}f_{(p)}-R_{(p)}R_{(p+1)})=\\
&=\fr\tr\left(\bar{X}X\pk1 \bar{X}X -X\bar{X}\k X\bar{X}\right)\label{we}
\end{split}\ee
\be\begin{split}
{\scriptstyle{(\ref{d})}}\quad \rightarrow\quad & R_{(k+1)}\sum_{p=1}^{k-1}(R_{(p)}^2-W^\prime_{(p)}R_{(p)}-\frac{1}{4}f_{(p)}-R_{(p)}R_{(p+1)})=\\
&=\fr\tr\left(XX\pk1 \bar{X}\bar{X} -X\bar{X}\k \bar{X}X\right)\label{wf}
\end{split}\ee
\be\begin{split}
{\scriptstyle{(\ref{e})}}\quad \rightarrow\quad & R_{(k)}\sum_{p=1}^{k}(R_{(p)}^2-W^\prime_{(p)}R_{(p)}-\frac{1}{4}f_{(p)}-R_{(p-1)}R_{(p)})=\\
&=\fr\tr\left(X\bar{X}\pk1 \bar{X}X-\bar{X}\bar{X}\k XX \right)\label{wg}
\end{split}\ee
\be\begin{split}
{\scriptstyle{(\ref{f})}}\quad \rightarrow\quad & R_{(k+1)}\sum_{p=1}^k(R_{(p)}^2-W^\prime_{(p)}R_{(p)}-\frac{1}{4}f_{(p)}-R_{(p-1)}R_{(p)})+\\&+R_{(k)}\sum_{p=1}^k(R_{(p)}^2-W^\prime_{(p)}R_{(p)}-\frac{1}{4}f_{(p)}-R_{(p)}R_{(p+1)})=\\
&=\fr\tr\left(\bar{X}X\pk1 \bar{X}X -X\bar{X}\k X\bar{X}\right)\label{wh}
\end{split}\ee
\be\begin{split}
{\scriptstyle{(\ref{g})}}\quad \rightarrow\quad & R_{(k+1)}\sum_{p=1}^{k-1}(R_{(p)}^2-W^\prime_{(p)}R_{(p)}-\frac{1}{4}f_{(p)}-R_{(p)}R_{(p+1)})=\\
&=\fr\tr\left(XX\pk1 \bar{X}\bar{X} -\bar{X}X\k X\bar{X}\right)\label{wi}
\end{split}\ee
\be\begin{split}
{\scriptstyle{(\ref{h})}}\quad \rightarrow\quad & R_{(k)}\sum_{p=1}^{k}(R_{(p)}^2-W^\prime_{(p)}R_{(p)}-\frac{1}{4}f_{(p)}-R_{(p-1)}R_{(p)})=\\
&=\fr\tr\left(\bar{X}X\pk1 X\bar{X}-\bar{X}\bar{X}\k XX \right)\label{wj}
\end{split}\ee
\end{subequations}
where
\be
\begin{split}
&f_{(k)}^{1,1}\equiv -\frac{1}{8\pi^2} \left(\frac{W_{(k)}^\prime(\Phi_{(k)})-W_{(k)}^\prime(z)}{z-\Phi_{(k)}}{\cal W}_\alpha^{(k)}{\cal W}^{(k)\alpha}X\bar{X}\right)\\
&f_{(k)}^{1,2}\equiv -\frac{1}{8\pi^2} \left(\frac{W_{(k)}^\prime(\Phi_{(k)})-W_{(k)}^\prime(z)}{z-\Phi_{(k)}}{\cal W}_\alpha^{(k)}{\cal W}^{(k)\alpha}\bar{X}X\right)
\end{split}
\ee
Given these Ward identities, all we need to do to write the cubic equation for
 the resolvents is to arrange a chain of variations (\ref{cubicvar}) so as to
 cancel all operators with four links. We start with the identity associated
 with (\ref{a}, $k=1$) which gives the operator 
$\tr X\bar{X}\frac{{\cal WW}}{z-\Phi_{(1)}} X\bar{X}$. If we add the identities
 for (\ref{c}, $k=1$), (\ref{b}, $k=2$), (\ref{d}, $k=2$), subtract the one
 for (\ref{g}, $k=2$) and add the relation coming from (\ref{a}, $k=2$) we end
 up again with an operator of the same form as the first one, but with 
$\Phi_{(2)}$ instead of $\Phi_{(1)}$. It is obvious now how we should go on:
 we add the same chain as before, $n-2$ more times, each time increasing all 
$k$'s by one, and end up with an expression in the $R_{(k)}$'s and
 $W^\prime_{(k)}$'s, with $X$ and $\bar{X}$ appearing only in
 $\tr X\bar{X}\frac{{\cal WW}}{z-\Phi_{(n-1)}} X\bar{X}$. At this point we add
 the equation for (\ref{c}, $k=n-1$) and are left with
 $\tr \bar{X}X\frac{{\cal WW}}{z-\Phi_{(n)}} \bar{X}X$, but this is exactly
 the only four-link operator appearing in (\ref{wc}), thus adding this last
 identity to our previous relation we find an equation that involves only
 $R_{(k)}(z)$, $W^\prime_{(k)}(z)$ and the polynomials $f(z)$
\be\label{cubicaAn}
\begin{split}
&\sum_{k=1}^{n-1}(R_{(k)} R_{(k+1)}^2-R_{(k)}^2 R_{(k+1)} )
+\sum_{k=2}^{n-1}W^\prime_{(k)}\left\{ \sum_{p=1}^{k}(R_{(p)}^2-W^\prime_{(p)}R_{(p)}-\frac{1}{4}f_{(p)})+\right.\\&\quad \left.-\sum_{p=k}^{n}(R_{(p)}^2-W^\prime_{(p)}R_{(p)}-\frac{1}{4}f_{(p)})-\sum_{p=2}^{k}R_{(p-1)}R_{(p)}+\sum_{p=k+1}^{n}R_{(p-1)}R_{(p)}\right\}+\\&\quad+W^\prime_{(1)}(R_{(1)}^2-W^\prime_{(1)}R_{(1)}-\frac{1}{4}f_{(1)})
-W^\prime_{(n)}(R_{(n)}^2-W^\prime_{(n)}R_{(n)}-\frac{1}{4}f_{(n)})+\\&\quad
+\frac{1}{4}\left(f_{(1)}^{1,1}+f_{(n)}^{1,2}+\sum_{k=2}^{n-1}(f_{(k)}^{1,1}+f_{(k)}^{1,2})\right)=0
\end{split}
\ee
where the $z$ dependence is understood. This is the cubic equation we were looking for, which is obtained by the chain of reparameterizations

\be
\sum_{k=1}^{n-1}(\delta\Phi_{(k)}^a+\delta X_{k,k+1}^c+\delta\Phi_{k+1}^b)+\sum_{k=2}^{n-1}(\delta X_{k,k+1}^d-\delta\bar{X}_{k+1,k}^g)
\ee

It is to notice that while for the variations that entered this chain we had to use all allowed $k$'s, some other reparameterizations never appeared at all. This might seem to suggest that generalization to higher order equations would need a case-by-case analysis to decide which transformations are useful in the process, and which are not. Based on the experience gained in writing the general cubic equation, in the next subsection we will suggest which reparameterizations are necessary, and which are redundant.

With the case of $A_2$ in mind, we can look for a change of variables which cancels, in the quadratic equation (\ref{quad1}), the terms linear in the 
resolvent. The new variables are:
\begin{align}
a_1 &\equiv - \sum_{p=1}^n \left( 1 - \frac{p}{n+1} \right) q_p \nonumber \\
a_k &\equiv - \sum_{p=k}^n q_p + \sum_{p=1}^n \frac{p}{n+1} q_p
\end{align}
for $k=2,...,n+1$, and where the $q_p$ are defined, in terms of the resolvents and the Cartan matrix $C_{qr}$ associated to the quiver, as
\begin{equation}
q_p = W'_p - C_{pr} R_r.
\end{equation}

In terms of the $a_{(i)}$, the $n$ equations satisfied by the resolvents of $A_n$ should become very simple   
if the curve underlying the quiver has the form suggested in \cite{Dijkgraaf:2002vw}:
\be\label{curvan}
(y-a_{(1)})(y-a_{(2)})...(y-a_{(n+1)})=0.
\end{equation}
As a consequence, the equation of order $k$ $(k=3,...,n+1)$ should be
\begin{equation}
\sum_{i_1<...<i_k}^{n+1} a_{(i_1)}...a_{(i_k)} = 
\mathrm{polynomial,~of~degree~}k \mathrm{,~built~with~} W_{(i)} 
\mathrm{~and~} f_{(i)} 
\end{equation}  
We checked this is the form of the cubic equation (\ref{cubicaAn}) in the new variables, which is an indication that the curve (\ref{curvan}) is correct.

\vspace{12pt}
{\it Vacuum expectation values of $4$-link operators}
\vspace{12pt}

The procedure we have just described also allows us to obtain expectation values for trace operators with four links. In fact at any intermediate step of the previous chain only one of these operators appeared at a time. As our previous observation already pointed out, not all of four-link operators can be reached this way, nonetheless there is a way to evaluate also their expectation values: this is where the Ward identities that didn't play a role in the anomaly chain, become necessary, as we will show.

For the operators in the chain, the expectation values are given by\footnote{Obviously the first identity in all expressions is not formally correct. It should be interpreted as representing the Ward identities we have to sum up to calculate the expectation value of the operator.}

\begin{subequations}
\be
\begin{split}
&\fr\tr X\bar{X}\k X\bar{X}=\\&\qquad =\sum_{p=1}^{k-1}(\delta\Phi_{(p)}^a+\delta X_{p,p+1}^c+\delta\Phi_{(p+1)}^b)+\sum_{p=2}^{k}(\delta X_{p,p+1}^d-\delta\bar{X}_{p+1,p}^g)+\delta\Phi_{(k)}^a=
\end{split}\nonumber
\ee\be\label{Xa}
\begin{split}
&\qquad =\sum_{p=1}^k (R_{(p)}^2 R_{p-1}-R_{p}^2 R_{(p+1)})-R_{(k)}\sum_{p=1}^k (R_{(p)}^2-W^\prime_{(p)}R_{(p)}-\frac{1}{4}f_{(p)}-R_{(p)}R_{(p+1)})+\\&\qquad +\sum_{p=1}^k W^\prime_{(p)}\left( \sum_{q=1}^p (R_{(q)}^2-W^\prime_{(q)}R_{(q)}-\frac{1}{4}f_{(q)}-R_{(q-1)}R_{(q)}+\right.\\&\qquad \left.-\sum_{q=p}^n (R_{(q)}^2-W^\prime_{(q)}R_{(q)}-\frac{1}{4}f_{(q)}-R_{(q)}R_{(q+1)}\right)+\frac{1}{4}\sum_{p=1}^k f_{(p)}^{1,1}+\frac{1}{4}\sum_{p=2}^k f_{(p)}^{1,2}
\end{split}
\ee
\be\label{Xb}
\begin{split}
&\fr\tr \bar{X}X\pk1 \bar{X}X=\\&\qquad =\sum_{p=1}^{k-1}(\delta\Phi_{(p)}^a+\delta X_{p,p+1}^c+\delta\Phi_{(p+1)}^b)+\sum_{p=2}^{k}(\delta X_{p,p+1}^d-\delta\bar{X}_{p+1,p}^g)+\delta\Phi_{(k)}^a+\delta X_{k,k+1}^d=\\
&\qquad=\fr\tr X\bar{X}\k X\bar{X}+R_{(k+1)}\sum_{p=1}^{k}(R_{(p)}^2-W^\prime_{(p)}R_{(p)}-\frac{1}{4}f_{(p)}-R_{(p-1)}R_{(p)})+\\&\qquad +R_{(k)}\sum_{p=1}^{k}(R_{(p)}^2-W^\prime_{(p)}R_{(p)}-\frac{1}{4}f_{(p)}-R_{(p)}R_{(p+1)})
\end{split}
\ee
\be\label{Xc}
\begin{split}
&\fr\tr X\bar{X}\pk1 \bar{X}X=\\&\qquad =\sum_{p=1}^{k}(\delta\Phi_{(p)}^a+\delta X_{p,p+1}^c+\delta\Phi_{(p+1)}^b)+\sum_{p=2}^{k}(\delta X_{p,p+1}^d-\delta\bar{X}_{p+1,p}^g)=\\&\qquad = \fr\tr \bar{X}X\k \bar{X}X+(W^\prime_{(k+1)}-R^\prime_{(k+1)})\sum_{p=1}^{k}(R_{(p)}^2-W^\prime_{(p)}R_{(p)}-\frac{1}{4}f_{(p)}+\\&\qquad -R_{(p)}R_{(p+1)})+\frac{1}{4}f_{(k+1)}^{1,2}
\end{split}
\ee
\be\label{Xd}
\begin{split}
&\fr\tr XX\frac{{\cal W}_\alpha^{(k+2)}{\cal W}^{(k+2)\alpha}}{z-\Phi_{(k+2)}} \bar{X}\bar{X}=\\&\qquad =\sum_{p=1}^{k}(\delta\Phi_{(p)}^a+\delta X_{p,p+1}^c+\delta\Phi_{(p+1)}^b)+\sum_{p=2}^{k}(\delta X_{p,p+1}^d-\delta\bar{X}_{p+1,p}^g)+\delta X_{k+1,k+2}^d=\\&\qquad =\fr\tr X\bar{X}\pk1 \bar{X}X\!+\!R_{(k+2)}\sum_{p=1}^{k}(R_{(p)}^2\!-\!W^\prime_{(p)}R_{(p)}-\frac{1}{4}f_{(p)}-\!R_{(p)}R_{(p+1)})
\end{split}
\ee
\be\label{Xe}
\begin{split}
&\fr\tr \bar{X}X\pk1 X\bar{X}=\\&\qquad =\sum_{p=1}^{k}(\delta\Phi_{(p)}^a+\delta X_{p,p+1}^c+\delta\Phi_{(p+1)}^b)+\sum_{p=2}^{k+1}(\delta X_{p,p+1}^d-\delta\bar{X}_{p+1,p}^g)=\\&\qquad =\fr\tr XX\frac{{\cal W}_\alpha^{(k+2)}{\cal W}^{(k+2)\alpha}}{z-\Phi_{(k+2)}} \bar{X}\bar{X}\!-\!R_{(k+2)}\sum_{p=1}^{k}(R_{(p)}^2\!-\!W^\prime_{(p)}R_{(p)}-\frac{1}{4}f_{(p)}-\!R_{(p)}R_{(p+1)})\\&\qquad =\fr\tr X\bar{X}\pk1 \bar{X}X
\end{split}
\ee
\end{subequations}

As can be seen from (\ref{Xe}), $\tr X\bar{X}\frac{\cal WW}{z-\Phi_{(k)}} \bar{X}X$ and $\tr\bar{X}X \frac{\cal WW}{z-\Phi_{(k)}} X\bar{X}$ have the same expectation values. This could already be seen by taking the difference of equations (\ref{wf}) and (\ref{wi}).
There is only one four-link operator whose expectation value could not be determined through the reparameterization that appeared in the chain. This can be determined, though, using (\ref{wg}) or (\ref{wj}) which by (\ref{Xe}) are equivalent. We have thus used all variations but two: (\ref{f}) and one of (\ref{e}) or (\ref{h}). As we will demonstrate in the next subsection, it is a general fact that variations like (\ref{c}) and (\ref{f}), which we will call alternating, give the same Ward identity.

\vspace{12pt}
{\noindent \it Outlook of the general procedure}
\vspace{12pt}

Based on the experience we gained in writing the general form of the quadratic and cubic equations for the resolvents, in this section we give some hints on how to write all the equations up to the $(n+1)^{\mathrm{th}}$ grade for a $A_n$ quiver theory, even though we won't be able to evaluate them explicitly.

First of all we want to give a rule that tells us which are the relevant reparameterizations we have to use.

It is a general feature of alternating perturbations, that is perturbations whose chain is an alternating array of $X$'s and $\bar{X}$'s, to give the same Ward identities. In fact, let us consider a generic degree $p$ alternating chain
\be
\begin{split}
&\delta X_{k,k+1}=\k X(\bar{X}X)^p\pk1 \\
&\delta \bar{X}_{k+1,k}=\pk1 \bar{X}(X\bar{X})^p\k 
\end{split}
\ee
then it is easy to show that both variations lead to the same Ward identity
\be
\begin{split}
-&\tr\pk1(\bar{X}X)^{p+1}+\tr \k(X\bar{X})^{p+1}=\\&=-\fr\sum_{m=0}^{p}\tr\k(X\bar{X})^m\cdot \tr\pk1(\bar{X}X)^{p-m}
\end{split}
\ee
One of these variations is then redundant, and must be neglected.

Let us now define a conjugation relation among $\delta X$ and $\delta \bar{X}$
 transformations: we say that a $\delta X$ and a $\delta \bar{X}$ variations 
are conjugated if the $X$-$\bar{X}$ chain of either can be obtained from the 
other's by inverting it and taking the ``complex conjugate'' (e.g. (\ref{d})
 and (\ref{g}) are conjugated in this sense). A $\delta X_{k,k+1}$ variation will contain as many $\bar{X}_{k+1,k}$ factors as ${X}_{k,k+1}$  are present in the $\delta\bar{X}_{k+1,k}$ conjugated variation, and, moreover, they are in specular positions. Thus the right hand side of the Ward identities (\ref{konishi}) generated by this pair will contain the same number of elements, and the number of links in each of the two factors of each element will be equal.

There is even more to this: the right hand side of the two Ward identities will be equal. In fact let us consider an $X_{k,k+1}$ field in the $\delta X_{k,k+1}$ variation, on both sides of it there will be an equal number of $X$ and $\bar{X}$ fields (even though the number of pairs on one side will in general be different from the number on the other side), because we are left with two fields in the adjoint one of $U(N_k)$ and the other of $U(N_{k+1})$. Let us now consider its image $\bar{X}_{k+1,k}$ through the conjugation operation. The same will be true for this field also, with representations exchanged. Now, because of the way we have defined conjugation, the subarray on the left of $X_{k,k+1}$ will be equal to the subarray on the right of $\bar{X}_{k+1,k}$ and viceversa. Let us make an example: we consider a general variation $\delta X_{k,k+1}$ and its conjugated partner and indicate with a hat the two conjugated links
\be
\begin{split}
&\delta X_{k,k+1}=\k X_{k,k+1}X\bar{X}\bar{X}\hat{X}_{k,k+1}X\bar{X}\frac{1}{z-\Phi_{(k+1)}}\\
&\delta\bar{X}_{k+1,k}=\pk1 X\bar{X}\hat{\bar{X}}_{k+1,k}XX\bar{X}\bar{X}_{k+1,k}\frac{1}{z-\Phi_{(k)}}
\end{split}
\ee

Thus taking the difference of the equations we obtain from two conjugated variations, gives an identity for four-link operators.

Unfortunately, we are not able to give an {\it a priori} prescription which tells us which conjugated pairs will give the same four-link identity. This will have to be checked case by case.

Once all variations and Ward identities have been written, it is not hard to write a $(k+2)^{th}$ degree equation involving only the resolvents. Let us suppose we have written the $(k+1)^{th}$ degree equation, and have evaluated the expectation values of all operators with $k$ pairs of bifundamentals. Then we only have to sum all reparameterizations for the adjoint fields and alternating transformations for $X$ containing $k$ couples $X$-$\bar{X}$, use the $(2k+2)$-link identities we have obtained from conjugated pairs, and make use of lower-degree equations. What we obtain is the $(k+2)^{th}$-degree equation for the resolvents of a $A_n$ ($n>k+1$) quiver theory.


\section{Affine $\hat{A}_1$ quiver}
The techniques developed in the previous sections can be
applied also to  affine quiver gauge theories.

Again we start by considering the simple case of $\hat{A}_1$ with
gauge group $U(N_1) \times U(N_2)$. In this case the number of
bifundamental fields is exactly two times the number of bifundamentals in the
$A_2$ theory; we call them $X_i$, $\bar{X}_i$ $i=1,2$.

The condition (\ref{condWaff}) on the superpotential implies 
\be\label{Wtotnullo}
W_2(x)=-W_1(x)\equiv -W(x).
\ee 
and the superpotential  (\ref{superpotential}) may be written as
\be
W_{tree}=\tr\left(W(\Phi_{(1)})-W(\Phi_{(2)})+\bar{X}_1\Phi_{(1)}X_{1}-X_1\Phi_{(2)}\bar{X}_{1}+X_2\Phi_{(2)}\bar{X}_{2}-\bar{X}_2\Phi_{(1)}X_{2} \right)
\ee   

In \cite{DVdec} it is showed that the matrix model associated to the
$\hat{A}_1$ quiver only depends on the combination
\be
S = S_1 -S_2,
\ee
where $S_k = g_s M_k$ for $k=1,2$ and $M_k$ is the rank of matrices
$\Phi_k$, and consequently that the planar loop equation is
identical to that of the bosonic one matrix model
\cite{Dijkgraaf:2002fc}: 
\be 
y^2 - W'(x)^2 +f(x)=0.  
\ee

In the following we reobtain the same result directly via a field theory
computation, using the variations of matter fields and the anomaly
equations introduced in section \ref{A2}. Indeed we will prove that
the difference of the resolvents $R_{(1)}(z) - R_{(2)}(z)$ satisfies the
quadratic equation of the $U(N)$ gauge theory with an adjoint $\Phi$:
\be\label{quadaffine} 
W'(z)\left( R_{(1)}(z) - R_{(2)}(z) \right) +\frac{1}{4}f(z) = 
\left( R_{(1)}(z) - R_{(2)}(z) \right)^2.
\ee

The anomaly equations that come from $\delta \Phi_{(1)}$ and $\delta \Phi_{(2)}$ are the same as 
(\ref{eq:anomalia1}) with the condition (\ref{Wtotnullo}):
\begin{align}
\label{eq:anomaff1}
 W'(z)R_{(1)}(z) + \frac{1}{4}f_1(z) -\fr\tr\bar{X}_1\p1 X_1 + \frac{1}{32\pi^2} \tr  \bar{X}_2 
\frac{{\cal W}_\alpha^{(1)} 
{\cal W}^{{(1)} \alpha}}{z-\Phi_{(1)}} X_2  =& R_{(1)}^2(z) \nonumber \\  
  -W'(z)R_{(2)}(z) -\frac{1}{4}f_2(z) +\fr\tr X_1\frac{{\cal W}_\alpha^{(2)} 
{\cal W}^{{(2)} \alpha}}{z-\Phi_{(2)}} \bar{X}_1- \frac{1}{32\pi^2} \tr X_2 
\frac{{\cal W}_\alpha^{(2)} 
{\cal W}^{{(2)} \alpha}}{z-\Phi_{(2)}} \bar{X}_2  \!=& R_{(2)}^2(z) 
\end{align}
where $f_k(z)$ is defined as usual.

The equations from $\delta X_1$ and $\delta X_2$ are
\begin{align}
\label{eq:anomaffine}
- \frac{1}{32\pi^2} \tr
 \bar{X}_1 \frac{{\cal W}_\alpha^{(1)} 
{\cal W}^{{(1)} \alpha}}{z-\Phi_{(1)}} X_1 +\fr\tr X_1 \frac{{\cal W}_\alpha^{(2)} 
{\cal W}^{{(2)} \alpha}}{z-\Phi_{(2)}} \bar{X}_1  =& R_{(1)}(z) R_{(2)}(z)\\
 \frac{1}{32\pi^2} \tr
 \bar{X}_2 \frac{{\cal W}_\alpha^{(1)} 
{\cal W}^{{(1)} \alpha}}{z-\Phi_{(1)}} X_2 -\fr\tr X_2 \frac{{\cal W}_\alpha^{(2)} 
{\cal W}^{{(2)} \alpha}}{z-\Phi_{(2)}} \bar{X}_2  =& R_{(1)}(z) R_{(2)}(z).
\end{align}

Summing the two equations (\ref{eq:anomaff1}) and subtracting both (\ref{eq:anomaffine}), we obtain, as expected, equation (\ref{quadaffine}) with $f(z)=f_1(z)-f_2(z)$.

To write the cubic equation we consider the variations
\begin{subequations}
\be
\delta\Phi_{(1),i}=\p1 X_i\bar{X}_i
\ee
\be
\delta\Phi_{(2),i}=\pW2 \bar{X}_i X_i
\ee
\be
\delta X_1=\p1 X_1 \bar{X}_1 X_1 \pW2
\ee
\be
\delta X_2=\p1 X_2 \bar{X}_2 X_2 \pW2
\ee
\be
\delta X_1=\p1 X_2 \bar{X}_2 X_1 \pW2
\ee
\be
\delta \bar{X}_1=\pW2 \bar{X}_1 X_2 \bar{X}_2 \p1
\ee
\be
\delta X_1=\p1 X_1 \bar{X}_2 X_2 \pW2
\ee
\be
\delta \bar{X}_1=\pW2 \bar{X}_2 X_2 \bar{X}_1 \p1
\ee
\end{subequations}

We do not write explicitly the Ward identities for these variations, since they are very similar to the equations we wrote for the non-affine $A_2$ theory, and moreover in the next section we write the explicit form of the cubic equation for generic $n$.

When we combine these Ward identities to give a cubic equation, we find that all resolvent dependences cancel, and we are eventually left with an equation for the polynomials $g_{(k,i)}(z)$
\begin{equation}
g_{(1,1)}(z)+g_{(1,2)}(z)+g_{(2,1)}(z)+g_{(2,2)}(z)=0
\end{equation}
where
\be\begin{split}
&g_{(1,i)}(z)\equiv \fr\tr\left(\frac{W^\prime_{(1)}(\Phi_{(1)})-W^\prime_{(1)}(z)}{z-\Phi_{(1)}} {\cal W}^{(1)}_\alpha {\cal W}^{(1)\alpha}X_i\bar{X}_i \right)\\
&g_{(2,i)}(z)\equiv \fr\tr\left(\frac{W^\prime_{(2)}(\Phi_{(2)})-W^\prime_{(2)}(z)}{z-\Phi_{(2)}} {\cal W}^{(2)}_\alpha {\cal W}^{(2)\alpha}\bar{X}_i X_i \right)
\end{split}
\ee

We believe all higher-degree equations will give no new constraints on the resolvents, just as we have showed the cubic doesn't.
We interpret this as saying that the low energy dynamics of the affine quiver $\hat{A}_1$ depends only on one resolvent (the difference of the two original ones) satisfying equation (\ref{quadaffine}), which is the generalized Konishi anomaly equation for a $U(N_1-N_2)$ gauge theory with tree superpotential $W(\Phi)$. This is in accordance with \cite{KS}, where this same theory has been studied from a much different point of view. There it was shown that under an RG flow the theory goes through a Seiberg-duality cascade, going all the way down to a $U(N_1-N_2+p)\times U(p)$ (where $p$ can also be 0). The same conclusion can be reached also from the study of this theory via matrix models, as has been done in \cite{DVdec}.


\newsection{Affine quivers $\hat{A}_{n-1}$}
{\it Quadratic equation}
\vspace{12pt}

Extending our previous analysis of generalized Konishi anomalies and resolvent equations to general affine quivers requires only little modifications. Let us start from the quadratic equation. We consider the same reparameterizations of the adjoint and bifundamental fields as in (\ref{adj}) and (\ref{bifund}), where this time $k=1,\ldots n$.
The Ward identities we obtain are
\be\label{affquadadj}
W^\prime_{(k)}(z)R_{(k)}(z)+\frac{1}{4}f_{(k)}(z)-\frac{1}{32\pi^2} \tr\bar{X} \k X+\fr\tr X \k\bar{X}  =R_{(k)}(z)^2
\ee
\be\label{affquadbif}
-\fr\tr\bar{X}\k X+\fr\tr X\pk1\bar{X}=R_{(k)}(z)R_{(k+1)}(z)
\ee
where again $f_{(k)}(z)$ is an $(n_k-1)^{th}$ degree polynomial defined as in (\ref{fz})
\be\label{fzaff}
f_{(k)}(z)\equiv -\frac{1}{8\pi^2}\tr\left(  \frac{ W^\prime_{(k)}(\Phi_{(k)})-W^\prime_{(k)}(z)}{z-\Phi_{(k)}}{\cal W}_{\alpha}^{(k)}{\cal W}^{(k)\alpha}\right)
\ee

We want to take a linear combination of these equations in order to eliminate all $X$- and $\bar{X}$-dependent operators. We find there is actually such a combination: we sum (\ref{affquadadj}) over $k=1,\ldots n$ and subtract (\ref{affquadbif}) summed over the same indices. The quadratic equation we obtain is
\be\label{affquad}
\sum_{k=1}^n(R_{(k)}^2(z)-W^\prime_{(k)}(z)R_{(k)}(z)-\frac{1}{4}f_{(k)}(z)-R_{(k)}(z)R_{(k+1)}(z))=0
\ee

Differently from the case of non-affine quivers, we can't write the expectation value of $X$-$\bar{X}$ operators as functions of the resolvents at this stage: the equations we have determined so far will prove not to be enough. In fact we can take the sum of (\ref{affquadadj}) and (\ref{affquadbif}) and find
\be\label{rec}
-\fr\tr\left( X\pk1 \bar{X}- X\k\bar{X}\right)=R_{(k)}^2-W^\prime_{(k)}R_{(k)}-\frac{1}{4}f_{(k)}-R_{(k)}R_{(k+1)}
\ee
Since we have no means to determine the value of any of the $X$-$\bar{X}$ operators, this equation needs a free-parameter to be introduced in order to give a solution. Let us call
\be
A\equiv -\fr\,\tr X\p1\bar{X}
\ee
then
\be\label{XbarX}
-\fr\tr X\pk1\bar{X}=A+\sum_{p=1}^k(R_{(p)}^2-W^\prime_{(p)}R_{(p)}-\frac{1}{4}f_{(p)}-R_{(p)}R_{(p+1)})
\ee

We can check that these identities are consistent with the quadratic equation (\ref{affquad}):
\be
-\fr\tr X\p1\bar{X}=-\fr\tr X\frac{{\cal WW}}{z-\Phi_{(n+1)}}\bar{X}=A
\ee

From (\ref{affquadbif}) we also determine
\be\label{barXX}
-\fr\tr\bar{X}\k X=A+\sum_{p=1}^k(R_{(p)}^2-W^\prime_{(p)}R_{(p)}-\frac{1}{4}f_{(p)}-R_{(p)}R_{(p+1)}) +R_{(k)}R_{(k+1)}
\ee

\vspace{12pt}
{\it Cubic equation}
\vspace{12pt}

To obtain the cubic equation we proceed in a similar manner. Let us consider the same transformations which we used to build the anomaly chain for the cubic equation of $A_n$ theories ($k=1,\ldots n$ this time) 

\begin{subequations}\label{affcubicvar}
\be
\label{affa} \delta\Phi_{(k)}=\k X\bar{X}
\ee\be
\label{affb} \delta\Phi_{(k)}=\k \bar{X}X
\ee\be
\label{affc} \delta X_{k,k+1}=\k X\bar{X}X\frac{1}{z-\Phi_{(k+1)}}
\ee\be
\label{affd} \delta X_{k,k+1}=\k\bar{X}XX\frac{1}{z-\Phi_{(k+1)}}
\ee\be
\label{affe} \delta \bar{X}_{k+1,k}=\pk1 \bar{X}\bar{X}X\frac{1}{z-\Phi_{(k)}}
\ee
\end{subequations}
from which we obtain

\begin{subequations}\label{affcubicward}
\be\label{affwa}
\begin{split}
(W^\prime_{(k)}-R_{(k)})\tr\bar{X}&\k X+\frac{1}{4}g_{(1,k)}=\\&=-\tr X\bar{X}\k X\bar{X} +\tr \bar{X}X\k X\bar{X}
\end{split}\ee
\be\label{affwb}\begin{split}
(W^\prime_{(k)}-R_{(k)})\tr X &\k \bar{X}+\frac{1}{4}g_{(2,k)}=\\&=-\tr X\bar{X}\k \bar{X}X +\tr \bar{X}X\k \bar{X}X
\end{split}\ee
\be\label{affwc}\begin{split}
R_{(k+1)}\tr\bar{X}\k & X +R_{(k)}\tr X\pk1 \bar{X}=
\\&=\tr X\bar{X}\k X\bar{X} -\tr\bar{X}X\pk1 \bar{X}X
\end{split}\ee
\be\label{affwd}
R_{(k+1)}\tr X\k \bar{X}=-\tr XX\pk1 \bar{X}\bar{X} +\tr X\bar{X}\k \bar{X}X
\ee
\be\label{affwe}
R_{(k+1)}\tr X\k \bar{X}=-\tr XX\pk1 \bar{X}\bar{X} +\tr\bar{X}X\k X\bar{X}
\ee
\end{subequations}
where
\be
\begin{split}
&\frac{1}{4}g_{(1,k)}(z)=\tr\left(\frac{W^\prime_{(k)}(\Phi_{(k)})-W^\prime_{(k)}(z)}{z-\Phi_{(k)}} {\cal W}_{\alpha}^{(k)}{\cal W}^{(k)\alpha}X\bar{X}\right)\\
&\frac{1}{4}g_{(2,k)}(z)=\tr\left(\frac{W^\prime_{(k)}(\Phi_{(k)})-W^\prime_{(k)}(z)}{z-\Phi_{(k)}} {\cal W}_{\alpha}^{(k)}{\cal W}^{(k)\alpha}\bar{X}X\right)
\end{split}
\ee

To build the cubic equation, we consider the chain
\be\label{affcubicchain}
\sum_{p=0}^{n-1}(\delta\Phi^a_{k+p}+\delta X^c_{k+p,k+1+p}+\delta\Phi^a_{k+1+p}+\delta X^d_{k+1+p,k+2+p}-\delta \bar{X}^e_{k+2+p,k+1+p})
\ee
where $k$ is any of the points on the quiver diagram. This chain allows us to take all 4-link operators out of the final Ward identity. The equation we obtain still contains 2-link traces, which we get rid of by using (\ref{XbarX}) and (\ref{barXX}). Finally, also $A$ does not appear because of (\ref{condWaff}). Fixing $k=1$ for simplicity, we obtain
\be
\begin{split}
\sum_{p=1}^n &W^\prime_{(p+1)}(z)\left(\sum_{q=1}^p(R_{(q)}^2(z)-W^\prime_{(q)}(z)R_{(q)}(z)-\frac{1}{4}f_{(q)}(z)-R_{(q)}(z)R_{(q+1)}(z))+\right.\\&\left. - \sum_{q=p+1}^n(R_{(q)}^2(z)-W^\prime_{(q)}(z)R_{(q)}(z)-\frac{1}{4}f_{(q)}(z)-R_{(q)}(z)R_{(q+1)}(z))\right)+\\&+\sum_{p=1}^n W^\prime_{(p)}(z)(R_{(p)}^2(z)-W^\prime_{(p)}(z)R_{(p)}(z)-\frac{1}{4}f_{(p)}(z))-\sum_{p=1}^n R_{(p)}^2(z) R_{(p+1)}(z)+\\&+\sum_{p=1}^n R_{(p-1)}(z) R_{(p)}^2(z)+\sum_{p+1}^n\left(\frac{1}{4}g_{(1,p)}(z)+\frac{1}{4}g_{(2,p)}(z)\right)=0
\end{split}
\ee

\newsection{Conclusions}
We have found the quadratic and cubic constraints on the resolvents of affine and non-affine quiver gauge theories. This allowed us to study in detail the two simplest examples $A_2$ and $\hat{A}_1$.

We have also given a procedure that allows to write all higher-order equations. We believe that the highest-degree equation for non-affine $A_n$ theories is of order $n+1$. This would mean that the resolvents are roots of an $(n+1)^{th}$ degree algebraic curve. Since we never worried of minimizing the effective superpotential, our analysis and the algebraic curve obtained from the anomaly equations are valid off-shell. The curve should be strongly related to the Seiberg-Witten curve of the theory \cite{Klemm} \cite{mtheorywitten} that is obtained going on-shell and turning off the tree superpotential to recover ${\cal N}=2$ supersymmetry. It still remains an open problem to make an on-shell analysis of these theories similar to what has been done for $U(N)$ theories in \cite{CSW2}.

In the affine case, it is not even clear what the algebraic curve should look like. There are some suggestions that it might be much more complicated than a finite order polynomial. Some more questions still wait for an answer in this case. First of all, it would be really interesting to understand how our picture changes when the two gauge factors of $\hat{A}_1$ are identical. Analysis undergone with different methods say that the theory becomes conformal in this case, and we believe this can be implemented in the formalism of this work only after having gone on-shell. Another non-trivial question to be addressed is to study the analog of RG cascades for $n>2$ affine quivers. There are examples of theories which are analogous to these models, for which such cascades exist \cite{cascade} \cite{tatar}.

We will study these open problems in a subsequent paper.

\vspace{12pt}
\noindent
{\bf Acknowledgments}
\vspace{12pt}

\noindent We would like to thank Alberto Zaffaroni for suggesting the problem under study and for useful discussions. We would also like to thank Luciano Girardello for discussions and suggestions. This work was partially supported by INFN and MURST under contract 2001-025492, and by the European Commission TMR program HPRN-CT-2000-00131, in association to the University of Padova.

\end{document}